# The Design of Sparse Antenna Array

Lianlin Li, wenji zhang and Fang Li
*Institute of Electronics, Chinese Academics of Sciences, Beijing, China*

*Abstract*- **The aim of antenna array synthesis is to achieve a desired radiation pattern with the minimum number of antenna elements. In this paper the antenna synthesis problem is studied from a totally new perspective. One of the key principles of compressive sensing is that the signal to be sensed should be sparse or compressible. This coincides with the requirement of minimum number of element in the antenna array synthesis problem. In this paper the antenna element of the array can be efficiently reduced via compressive sensing, which shows a great improvement to the existing antenna synthesis method. Moreover, the desired radiation pattern can be achieved in a very computation time which is even shorter than the existing method. Numerical examples are presented to show the high efficiency of the proposed method.**
*Index Terms*- **compressive sensing, antenna array, antenna synthesis**

## I. INTRODUCTION

The aim of antenna array synthesis is to achieve a desired radiation pattern with the minimum number of antenna elements. This is of particularly use in many applications where the weight and size of antennas are extremely limited, such as phased array radar, satellite communication [1, 2, 3].

Up to present many analytical formulations have been presented for the antenna problem, such as Dolph-Chebyshev and Taylor methods. This kinds of methods generally based on an assumption that the antenna element is equally spaced with a uniform distribution thus a large number of antenna element are obtained using these methods. Some deterministic and global optimization methods have been developed for synthesis the desired pattern with uniformly distributed antenna elements, such as genetic algorithm, particle swarm optimization method and differential evolution algorithm. A novel non-iterative algorithm based on the matrix pencil method has been proposed recently which can efficiently reducing the number of antennas in a very short computation time.

In this paper the antenna synthesis problem is studied from a totally new perspective: compressive sensing. One of the key principles of compressive sensing is that the signal to be sensed should be sparse or compressible [4,5,6]. This coincides with the requirement of minimum number of element in the antenna array synthesis problem.

The CS theory asserts that one can recover certain signals and images from far fewer samples or measurements than traditional methods use. To make this possible, compressive sampling relies on two principles: sparsity, which pertains to the signals of interest, and incoherence, which pertains to the sensing modality [4]. Compressive sampling exploits the fact that many natural signals are sparse or compressible in the sense that they have concise representations when expressed in the proper basis. In the antenna synthesis problem if the location, amplitude and phase distribution of all the elements in the array can be obtained in the sparsest representation the aim of synthesizing a desired radiation pattern with the minimum antenna elements is achieved.

## II. ANTENNA SYNTHESIS via COMPRESSIVE SENSING

The antenna synthesis problem can be described as follows. Suppose that the antenna array is composed of N identical antenna element. The antenna factor is given by

$$F(\theta) = \sum_{i}^{N} R_i e^{jkd_i \cos\theta} \quad (1)$$

where $R_i$ is the complex excitation coefficient of the *i*th element located at $x = d_i$, $k$ is the wave number in the freespace $k = \frac{2\pi}{\lambda}$. The aim of the antenna synthesis problem is to achieve a desired radiation pattern with the minimum number of antenna element. Suppose that all the antenna elements are symmetric distributed within a range of $-x_m$ to $x_m$ along the *x* direction (Fig.1).

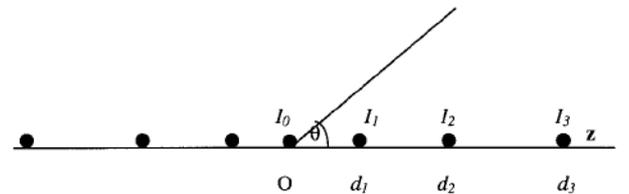

Fig1. Geometry of nonuniformly spaced linear symmetric array in [3]

The actual array pattern of the above symmetric array can be written as

$$F(\theta) = 2\sum_{i}^{[N/2]} R_i \cos(kd_i \cos\theta) \quad (2)$$

where $[N/2]$ is the minimum integer no less than N/2. In order to solve the above equation we can suppose that all the antenna elements are equally spaced from $-x_m$ to $x_m$ with a very small step $\Delta d$. Then equation (2) can be written in a matrix form

$$[f]_{m \times 1} = [A]_{m \times n} [r]_{n \times 1} \qquad (3)$$

where $m$ is the number of sampled antenna radiation pattern data, $n$ is the minimum integer no less than $\frac{2x_m}{\Delta d}$,

$$f = [F(\theta_1), F(\theta_2), \cdots F(\theta_N)]^T,$$
$$r = [R_1, R_2, \cdots R_n]^T$$
$$A_{ij} = 2R_i \cos(kd_i \cos \theta_j)$$

All of the existing algorithms do not exploit any prior information about the antenna location space with the current antenna synthesis framework. Although the antennas are supposed to be equally spaced with a small distance between each other, not every antenna needs to radiate the EM waves, which means that in fact there is no antenna element in the supposed location if it does not radiate waves. Thus we can exploit very useful priori information that the antenna location space is sparse. However, most of the existing algorithms don't use this prior knowledge of the antennas. Spatial sparsity means that the number of actual antennas is much less than that supposed to form Eq (3).

Then the antenna synthesis problem can be cast as a L1 convex minimization problem.

$$\min \|R\|_{l1}, \quad \text{s.t.} \ [f] = [A][r] \qquad (4)$$

In the traditional optimization method the above problem is always cast as L2 minimization problem and solved iteratively. As is all known the L2 constrain does not ensure a sparse representation of the signal. Even the iterative method for solving the L2 minimization problem can efficiently reduce the number of element in the array, it is very time consuming.

## III NUMERICAL RESULTS

In this section some numerical results are presented to show the effectiveness of the proposed method. The first example is to synthesis a twenty uniformly spaced Chebyshev array. The side-lobe level of the Chebyshev array is SLL=-30dB. In the simulation $x_m = 10\lambda$, $\Delta d = \lambda/10$. Table 1 gives the antenna location and excitation amplitude in [2] and obtain by the method in this paper. Fig. 2 is the reconstructed radiation pattern with compressive sensing and the methods in [1] and [2]. From the result in Fig.2 and Table.1 we can find that the antenna element is reduced 40% than the uniformly spaced array and the synthesized pattern is quite satisfactory for practical use.

The second example is chosen to be the same as Example2 in [2]. The ain is to synthesize the Taylor-Kaise array radiation pattern with the minimum antenna element. The position and amplitudes of the reconstructed nonuniformly array element with compressive sensing and those in [1] and [2] are also presented. Fig. 3 is the reconstructed radiation pattern with compressive sensing. From the reconstruction result we can find that by using compressive sensing only 18 antenna element is enough to achieve the desired radiation pattern. It saves about 38% antenna element than the uniformly space antenna array and although it is one more than the number in [2] is still very satisfactory.

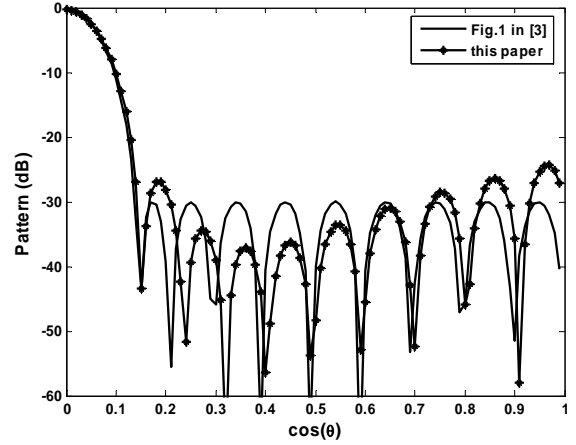

Fig. 2 Reconstruction of the 20 element Chebyshev pattern by nonuniform arrays with 12 antenna element via compressive sensing

| | Chebyshev (N=20) | | Position/excitation in [2] | | Position/excitation with CS | |
|---|---|---|---|---|---|---|
| i | $d_i/\lambda$ | $R_i$ | $d_i/\lambda$ | $R_i$ | $d_i/\lambda$ | $R_i$ |
| 1 | 0.25 | 0.25 | 0 | 1 | 0.4 | 1 |
| 2 | 0.75 | 0.75 | 0.8206 | 0.95818 | 1.2 | 0.92947 |
| 3 | 1.25 | 1.25 | 1.6381 | 0.84113 | 2 | 0.79514 |
| 4 | 1.75 | 1.75 | 2.4481 | 0.67176 | 2.8 | 0.61428 |
| 5 | 2.25 | 2.25 | 3.2432 | 0.48115 | 3.6 | 0.41719 |
| 6 | 2.75 | 2.75 | 4.0071 | 0.30046 | 4.5 | 0.2409 |
| 7 | 3.25 | 3.25 | 4.7145 | 0.23345 | | |
| 8 | 3.75 | 3.75 | | | | |
| 9 | 4.25 | 4.25 | | | | |
| 10 | 4.75 | 4.75 | | | | |

Table 1. Position and amplitudes of the reconstructed nonuniformly array elements for the Chebyshev radiation pattern

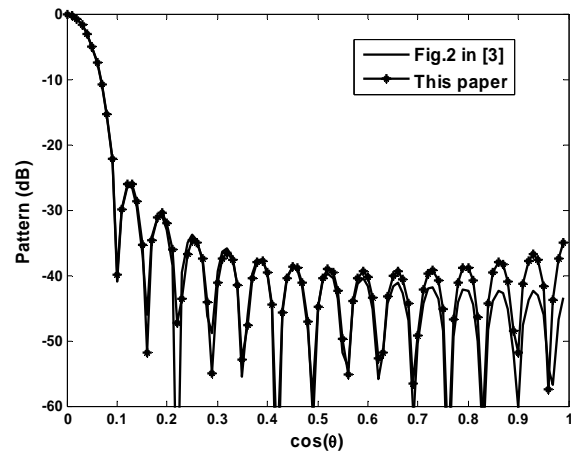

Fig. 3 Reconstruction of the 29 element Taylor-Kaise arrays with 18 antenna elements via compressive sensing

| i | Position/excitation in [1] | | Position/excitation in [2] | | Position/excitation with CS | |
|---|---|---|---|---|---|---|
| | $d_i/\lambda$ | $R_i$ | $d_i/\lambda$ | $R_i$ | $d_i/\lambda$ | $R_i$ |
| 1 | 0 | 1 | 0 | 1 | 0.4 | 1 |
| 2 | 0.5 | 0.99328 | 0.8831 | 0.97859 | 1.2 | 0.96431 |
| 3 | 1 | 0.97329 | 1.7652 | 0.91634 | 2 | 0.89689 |
| 4 | 1.5 | 0.94063 | 2.6451 | 0.81903 | 2.8 | 0.8045 |
| 5 | 2 | 0.89622 | 3.5211 | 0.69547 | 3.6 | 0.69429 |
| 6 | 2.5 | 0.84132 | 4.3905 | 0.55651 | 4.4 | 0.57146 |
| 7 | 3 | 0.77748 | 5.2485 | 0.4137 | 5.2 | 0.43998 |
| 8 | 3.5 | 0.70645 | 6.0842 | 0.27782 | 6 | 0.30694 |
| 9 | 4 | 0.63017 | | | 6.8 | 0.18562 |
| 10 | 4.5 | 0.55065 | | | | |
| 11 | 5 | 0.46994 | | | | |
| 12 | 5.5 | 0.39004 | | | | |
| 13 | 6 | 0.31282 | | | | |
| 14 | 6.5 | 0.24001 | | | | |
| 15 | 7 | 0.17309 | | | | |

Table 2. Position and amplitudes of the reconstructed nonuniformly array elements for the Taylor-Kaise radiation pattern

## IV. CONCLUSION

In this paper the antenna synthesis problem is studied from a totally new perspective. All of the existing algorithms do not exploit any prior information about the antenna location space within the current antenna synthesis framework. By exploiting the priori information that the antenna location space is sparse, the antenna element can be efficiently reduced via compressive sensing.

APPENDIX A. The description of Compressive sensing (CS)

CS relies on the surprising observation that a signal having a sparse representation in one basis (e.g. DCT, wavelet, curvelet, overcomplete atoms, and so on) can be recovered from a small number of projections onto a second basis that is incoherent with the first one. In particular, for an N-sample signal that is K-sparse or approximately K-sparse, roughly cK (cK<<N) projections are required for an exact reconstruction of the signal with high probability.

For the K-sparse signal $\theta$ (no more than K non-zero components), it has been demonstrated that one may recover $\theta$ *exactly* using a $l_p \, (p<2)$ constrained linear program under certain conditions, in particular,

$$\min \|x\|_{l_p}, \text{ s.t. } \Phi\theta = f. \quad \text{(A. 1)}$$

Moreover, it has been demonstrated that if the entries of the M by N sensing matrix $\Phi$ are draw i.i.d. from a normal distribution with mean zero and variance 1/N, then if

$$K \leq c \frac{M}{\log(N/M)} \quad \text{(A. 2)}$$

the K-sparse $\theta$ can be *exactly* reconstructed with "*overwhelming*" probability.

Considering the K-sparse signal and the complex matrix $\Phi$ with size of M by N (M~O(K)<<N), let $T \subset J$ be a set of indices, where the set $J = \{1, 2, \cdots, m\}$ indexes all columns of $\Phi$; $T$ identifies an indexed subset of columns in $\Phi$, and $\Phi_T$ represents the submatrix $\Phi$ for which only these columns are retained.

Consider $\Phi_T c = \sum_{j \in T} c_j \phi_j$, where $\phi_j$ represents the *j*th column of $\Phi$, and the $c_j$ are the arbitrary complex numbers. Candes and Tao define the K-restricted isometry constants $\delta_K$ to be the smallest quantity such that $\Phi_T$ obeys

$$(1-\delta_K)\|c\|_{l_2}^2 \leq \|\Phi_T c\|_{l_2}^2 \leq (1+\delta_K)\|c\|_{l_2}^2 \quad \text{(A. 3)}$$

for all subsets $T \subset J$ of cardinality $|T| \leq K$ and all complex coefficients $\{c_j\}_{j \in T}$.

Acknowledgement:
Dr. Lianlin Li would like to thank S. Ji, Y. Xue and L. Carin for their public Matlab Codes about Bayesian Compressive Sensing.
This work has been supported by the National Natural Science Foundation of China under Grants 60701010 and 40774093.

## REFERENCES

[1] S. J. Orfanidis, Electromagnetic Waves and Antenna 2004 [Online],
Available: http://www.ece.rutgers.edu/~orfanidi/ewa/.
[2] Yanhui Liu, Zaiping Nie, Qinghuo Liu, Reducing the number of elements in a linear anenna array by the matrix pencil method, IEEE Trans. on Antennas and propagation, 56 (9), 2955-2962, 2008
[3] B. P. Kumar, G. R. Branner, Design of Unequally Spaced Arrays for Performance Improvement, Trans. on Antennas and propagation, 47 (3), 511-523, 1999
[4]E. Candes, J. Romberg and T. Tao, Roubust uncertainty principles: exact signal reconstruction from highly incomplete frequency information, IEEE Trans. on Information Theory, 52(2), 489-509, 2006
[5]D. Donoho, Compressed sensing, IEEE Trans. on Information Theory, 52(4), 1289-1306, 2006
[6]S. Ji, Y. Xue and L. Carin, Bayesian compressive sensing, IEEE Trans. on Signal processing, 56(6), 2346-2356, 2008